\documentclass[lettersize,journal]{IEEEtran}
\usepackage{amsmath,amsfonts}
\usepackage{algorithmic}
\usepackage{algorithm}
\usepackage{array}
\usepackage{textcomp}
\usepackage{stfloats}
\usepackage{url}
\usepackage{verbatim}
\usepackage{graphicx}
\usepackage{cite}
\usepackage{caption}
\usepackage{amsmath,amssymb,amsfonts}
\usepackage{upgreek}
\usepackage{mwe}
\usepackage{color,soul}
\begin{document}

\title{Terahertz Bandstop Filter Using Varying Radii Split-Ring Resonators}

\author{Saeid~Asadi$^{1,2}$,
        Mohsen~Haghighat$^{1,2}$,
        Thomas~Darcie$^{1}$,~\IEEEmembership{Fellow,~IEEE}
        and Levi~Smith$^{1,2,*}$,~\IEEEmembership{Member,~IEEE}
\thanks{$^1$Department of Electrical and Computer Engineering, University of Victoria, Victoria, BC, V8P 5C2 Canada}
\thanks{$^2$Centre for Advanced Materials and Related Technology (CAMTEC), University of Victoria, 3800 Finnerty Rd, Victoria, BC, V8P 5C2, Canada.}
\thanks{$^*$Corresponding author: levismith@uvic.ca}
\thanks{This work was supported by a NSERC Discovery Grant.}
\thanks{This work made use of the 4D LABS at Simon Fraser University.}
\thanks{Manuscript received May, 2024; revised March x, xxxx.}}

\markboth{IEEE Template,~Vol.~x, No.~x, May~2024}%
{Asadi \MakeLowercase{\textit{et al.}}: A THz Bandstop Filter Using Split-Ring Resonators}


\maketitle

\begin{abstract}
In this letter we report a proof-of-concept terahertz band-stop filter constructed from split-ring resonators that has a center frequency of 1.06 THz and a -3 dB bandwidth of 0.36 THz. The design consists of nine split ring resonators of varying radii (3 $\times$ 13 \textmu m, 3 $\times$ 14 \textmu m, 3 $\times$ 15 \textmu m) that are placed between the conductors of a coplanar stripline transmission line. The response of the filter is measured using a modified terahertz time-domain spectrometer and we find reasonable agreement between simulation and experiment. This work demonstrates the viability of using varying-radii split-ring resonators as discrete sub-wavelength filter elements for terahertz systems.
\end{abstract}

\begin{IEEEkeywords}
Terahertz, guided wave, filter, coplanar strip line, band-stop filter, split ring resonator, resonant frequency, metamaterial. 
\end{IEEEkeywords}

\section{Introduction}

\IEEEPARstart{F}{ilters} are key elements that are used throughout the electromagnetic spectrum for many applications such as noise reduction \cite{oppenheim_signals_1997}, multiplexing \cite{brackett1990dense}, electromagnetic compatibility \cite{paul2022introduction}, and safety \cite{tuchinda2006photoprotection}. The desired filter configuration (low-pass, high-pass, band-pass, band-stop) is dependent on the specific application. Among filters, band-stop filters are used to eliminate spurious signals, which impact the performance of systems \cite{garcia2005microwave}. There are several common band-stop filters configurations which use gratings \cite{park1985bandstop} or resonant elements \cite{lee2021transmission}.  There are many resonators such as a hollow cavity in a metallic structure \cite{nickelson2019rectangular}, transmission line stubs \cite{zhu2003compact}, dielectric material such a whispering-gallery resonator \cite{vogt2018ultra}, or, as used in this work, split-ring resonators (SRRs) \cite{martin2003split,falcone2004babinet,horestani2013metamaterial}. Here SRRs are used due to their small size which can have diameters $<0.1 \lambda$ to enable compact filters when compared with Bragg filters which require element periods of $0.5 \lambda$.

This work focuses on the integration of  band-stop filters with transmission lines (TLs) frequencies near or above 1 THz. Under these conditions there are a limited number of viable TL options due to excessive loss or dispersion \cite{6005337, gallot2000terahertz}. One option which performs well is a coplanar-strip (CPS) TL on a thin Silicon Nitride membrane. The CPS TL geometry is used for its compatibility with photoconductive switching \cite{frankel1991terahertz}, lower attenuation and dispersion \cite{smith2019demonstration,cheng1994terahertz}, and its electromagnetic field properties (magnetic field is aligned with the SRR surface normal vector). A thin dielectric substrate is necessary to minimize dispersion and loss for the propagating wave while maintaining mechanical rigidity. Here, a thin silicon-nitride membrane substrate was selected for its compatibility with lithographic processing and its acceptable dielectric properties. In our prior work we have performed a similar experiment but focused on a single SRR as a proof-of-concept demonstration for a notch filter \cite{smith2021characterization}. This work expands on our prior work by experimentally demonstrating that bandstop filter (opposed to a notch filter) using similar methodology is viable option if the radii of the SRRs is varied.

We briefly review several examples of using SRRs as filters which are most commonly found at sub-THz frequencies. In \cite{martin2003split}, a notch filter was integrated into a coplanar waveguide (CPW) by placing double-SRRs on the backside of the substrate which demonstrated a center frequency around 8 GHz and -3dB bandwidth of 1.5 GHz. In \cite{ponchak2018coplanar}, a notch filter was fabricated by loading a CPS transmission line with double-SRR adjacent to the conductors which achieved a center frequency about 5 GHz and -3dB bandwidth of 0.5 GHz. In \cite{atallah2019design}, a wide band-stop, with rejection band about 3.56 GHz, is simulated by using different complimentary double-SRR on the ground plane of a microstrip (MS) TL. In \cite{naqui2013modeling}, notch filters with loading double-SRR and complimentary double-SRR to microstrip TLs working around 1.3 GHz are fabricated.  In \cite{fertas2017design}, double-SRRs with different sizes are fabricated alongside a microstrip TL to create three notch band-stop filters from 6 to 8 GHz. In \cite{shaterian2022multifunctional}, single-SRR are fabricated on another substrate over the substrate of microstrip TL and the result is a notch filter working around 4.9 GHz. Goubau TL notch filters about a few hundreds of GHz with single and double-SRR are reported in \cite{park2020determination} and \cite{parker2021tunable}, respectively. Also, as mentioned, we demonstrated a notch filter in \cite{smith2021characterization} where a double-SRR are loaded inside the CPS, and the center frequency of this structure is 500 GHz. These filters are summarized in Table \ref{tab:lit_rev}.

\begin{table}[!h]
\caption{Examples of filters made of SRRs}
\label{tab:lit_rev}
\centering
\begin{tabular}{|l|l|p{1cm}|l|p{1.2cm}|l|}
\hline
Ref. & Feedline & Resonator & $\it{f_c}$[GHz] & -3dB BW [GHz] & Rejection [dB]\\
\hline
\cite{martin2003split} & CPW & DSRR & 8 & 1.05 & -37\\
\hline
\cite{ponchak2018coplanar} & CPS & DSRR & 5.4 & 0.5 & -2.4\\
\hline
\cite{atallah2019design} & MS & CDSRR & 6.2 & 3.56 & -42, -53, -38\\
\hline
\cite{naqui2013modeling} & MS & DSRR, CDSRR & 1.3 &0.08, 0.18 &-40, -40\\
\hline
\cite{fertas2017design} & MS & DSRR & 6, 7, 8 & 0.1 & -8\\
\hline
\cite{shaterian2022multifunctional} & MS & SRR & 4.9 & 0.36 & -40\\
\hline
\cite{park2020determination} & Goubau & SRR & 200-700 & 45 & -30\\
\hline
\cite{parker2021tunable} & Goubau & DSRR & 220-300 & 60 & -30\\
\hline
\cite{smith2021characterization} & CPS & DSRR & 500 & 60 & -36\\
\hline
\multicolumn{6}{l}{* Values are obtained using digitized graphs} 
\end{tabular}
\end{table}

Many band-stop filters, mostly notch filters, have been studied in previous works. However, to our knowledge, there are no experimental results of band-stop filters which use varying radii SRRs at THz frequencies. This work demonstrates, via experiment and simulation, a wideband filter at THz frequencies which is constructed using a CPS TL loaded with nine single-SRRs of different radii.

\section{Background}
\begin{figure}[!h]
\centering
\includegraphics[width=3in]{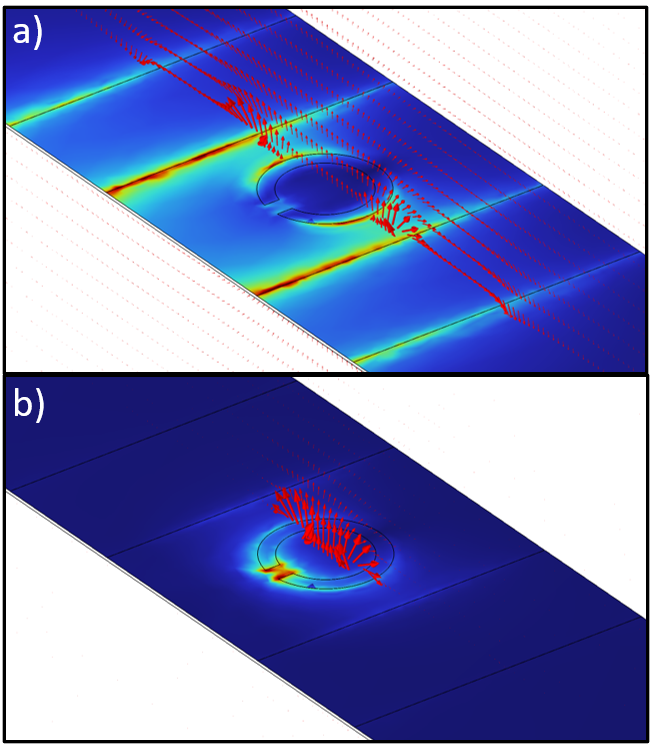}
\caption{Electromagnetic field for SRR in CPS TL. a) Off resonance (f = 0.5 THz). b) On resonance (f = 1 THz).  Red arrows: $\Vec{H}$. (COMSOL)}
\label{fig:em_field}
\end{figure}

Figure \ref{fig:em_field} illustrates the magnetic-field vector for a CPS TL containing a SRR which demonstrates the coupling between the CPS TL and SRR on and off resonance. The TL-coupled SRR can be thought of as a coupled inductor in series with a capacitor and resistor. The inductor represents the magnetic energy stored around SRR conductor. The series capacitor represents the electric energy stored in the gap of the SRR. The series resistor represents the ohmic losses due to the finite conductivity and a radiation resistance of a small loop antenna. When the inductive and capacitive reactances balance, then the incident signal is dissipated in the resistive element (heat and radiation) and the transmission line exhibits stop-band behavior \cite{martin2003split, syms2005theory, baena2005equivalent, smith2021characterization}. Each of these equivalent parameters (R, L, C) are dependent on the radius of the SRR. In this work, we slightly vary the radii of three different group of SRRs to obtain different resonant frequencies. This combination of different radii SRRs then converts a notch filter into a bandstop filter.

\begin{figure*}[!h]
\centering
\includegraphics[width=\linewidth]{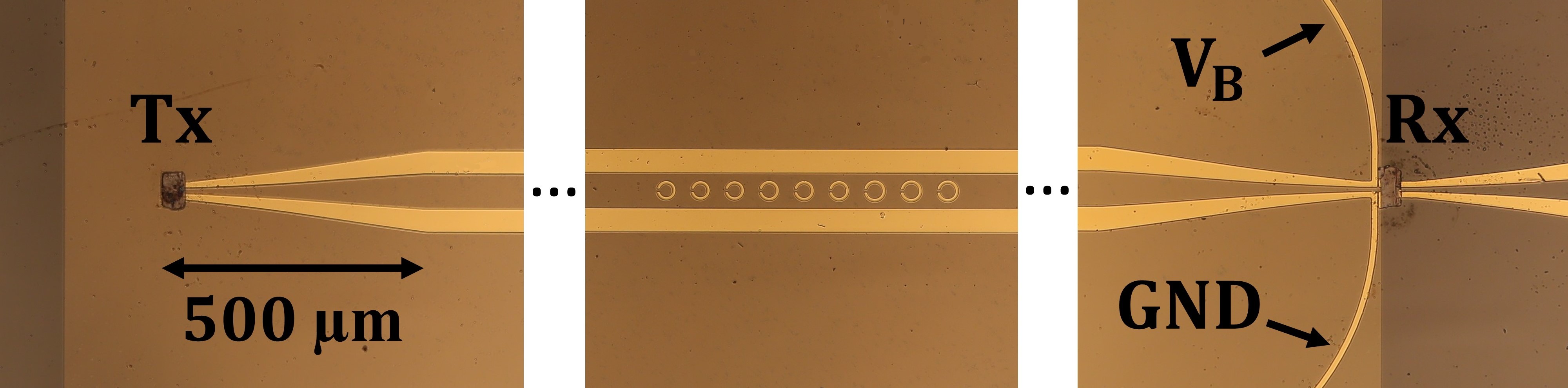}
\caption{CPS loaded with nine single-SRR made of gold. The inner radii of the three left, middle, and right rings are $13~ \upmu$m, $14 ~\upmu$m, and $15 ~\upmu$m, respectively.}
\label{fig:9srr_pic}
\end{figure*}

\section{Design and Simulation}

\begin{figure}[!h]
\centering
\includegraphics[scale=0.26]{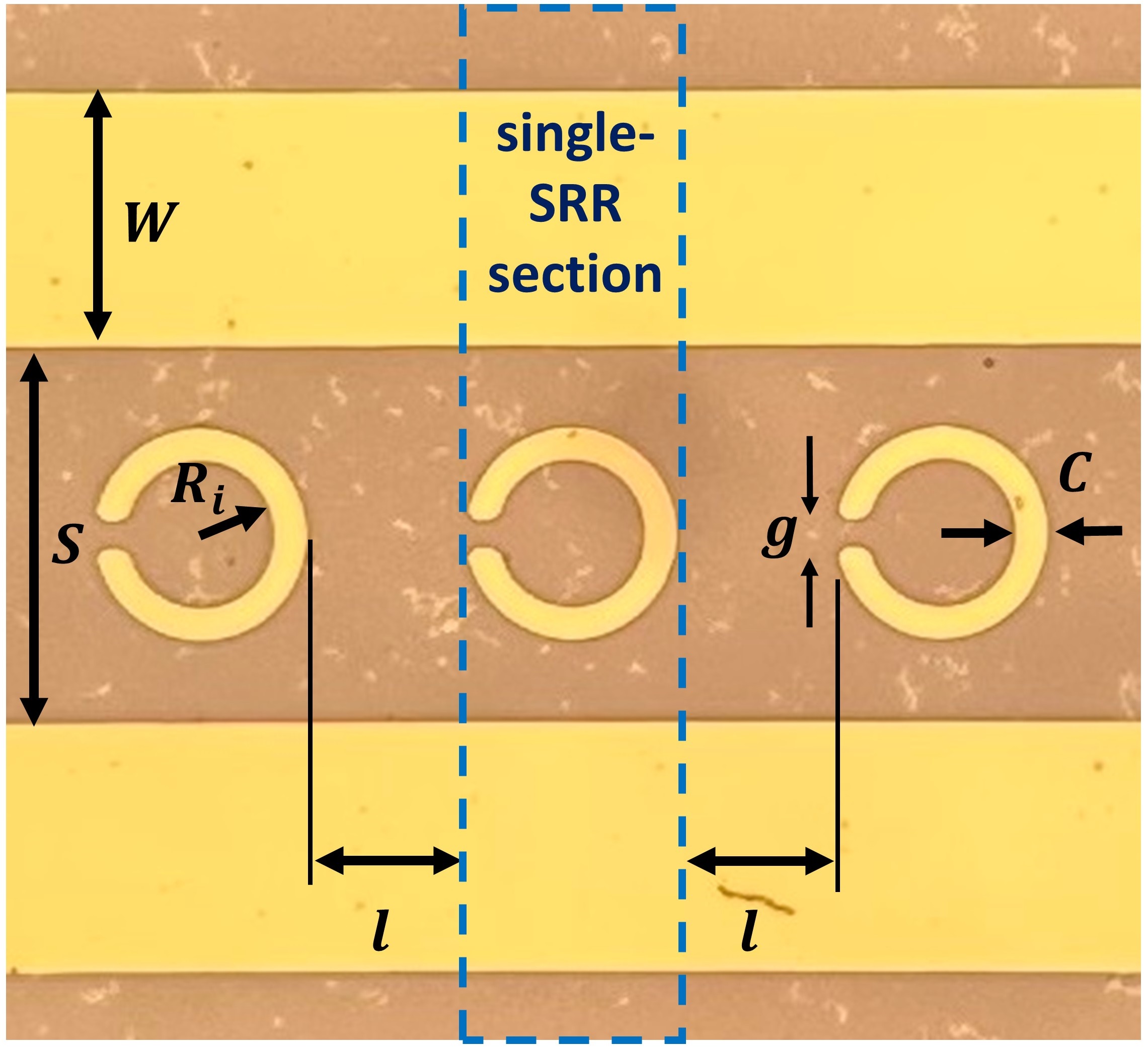}
\caption{A group of three single-SRR. $W = 45 ~\upmu$m, $S = 70~ \upmu$m, $l = 30 ~\upmu$m, $g = 6~ \upmu $m, and $C = 5 ~\upmu$m.}
\label{fig:3srr_pic}
\end{figure}

Figure \ref{fig:9srr_pic} illustrates the SRR BSF which consists of three groups of three single SRRs for a total of nine single SRRs. Each group of single SRRs has a different radius to achieve a different resonant frequency and each group contains three single SRRs to increase the stop-band rejection. Figure \ref{fig:3srr_pic} illustrates an annotated group of single SRRs where $W$ is the width of the CPS, $S$ is the distance between CPS, $l$ is the distance between rings, $R_i$ is the inner radius of rings, $g$ is the gap, and $C$ is thickness of the rings. For all groups, the dimensions are $W = 45 ~\upmu$m, $S = 70 ~\upmu$m, $l = 30~ \upmu$m, $g=6~ \upmu$m, and $C = 5~ \upmu$m.

\begin{figure}[!h]
\centering
\includegraphics[width=2.5in]{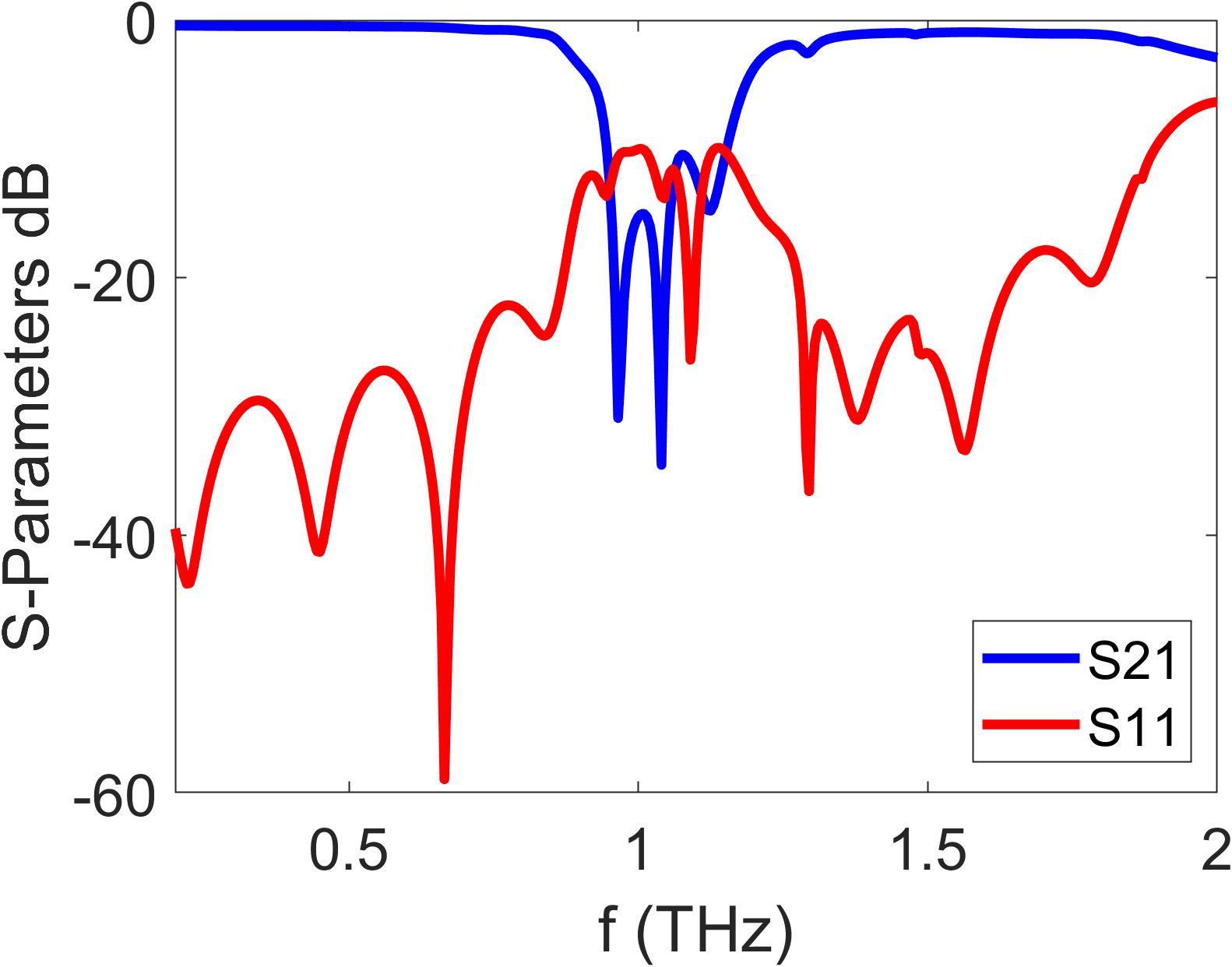}
\caption{S-parameters (simulation) for the CPS loaded with nine single-SRR with different radii of 13 $\upmu$m, 14 $\upmu$m, and 15 $\upmu$m.}
\label{fig:S21_9srr_sim}
\end{figure}

We aimed to design a band-stop filter centered near 1 THz with a -3 dB bandwidth exceeding $\approx$0.25 THz. It is necessary to select the radius of each SRR group to achieve these specifications. This task was performed using an Eigenmode simulation in ANSYS HFSS which resulted in the inner radius are $R_1 = 13~ \upmu$m ($f_{r1}$ = 1.12 THz), $R_2 = 14~ \upmu$m ($f_{r2}$ = 1.06  THz), and $R_3 = 15~ \upmu$m ($f_{r3}$ = 0.975 THz). For all simulations, the material properties are $\sigma$\textsubscript{Au} $=41$ (MS ⁄ m), $\epsilon_r$(Si$_3$N$_4$) $=7.6$, and the loss tangent of Si$_3$N$_4$ is 0.00526 \cite{Cataldo_Silicon_nitride_properties_2012}. 
The simulated scattering parameters for the total structure illustrated in Fig. \ref{fig:9srr_pic} are shown in Fig. \ref{fig:S21_9srr_sim}. The center frequency is located at 1.06 THz and the -3 dB bandwidth is 0.32 THz. From $|\text{S}_{21}|$ it is clear that the combined structure exhibits bandstop behavior around 1 THz. Also, since $|\text{S}_{21}| + |\text{S}_{11}| < 1$, the filter exhibits loss (heat and radiation) in the stopband which may be desirable in applications where the transmitter is sensitive to reflected power.

The transmitter (Tx) and receiver (Rx) are thin photoconductive switches made of low-temperature-grown GaAs (LT-GaAs) which are Van der Waals bonded to the surface and CPS TL (Fig. \ref{fig:9srr_pic}) \cite{rios2015bow}. Their dimensions are $20 ~\upmu $m$ \times 40 ~\upmu $m$ \times 1.8~ \upmu $m. The distance between Tx and Rx is 10 mm. As previously mentioned, to enable THz-bandwidth pulse transmission, a suspended thin (1 \textmu m) Si$_{3}$N$_{4}$ is used as the substrate. If, instead, a thick substrate was used then significant substrate radiation would dominate the loss mechanism \cite{cheng1994terahertz}.

\begin{figure}[!h]
\centering
\includegraphics[width=\linewidth]{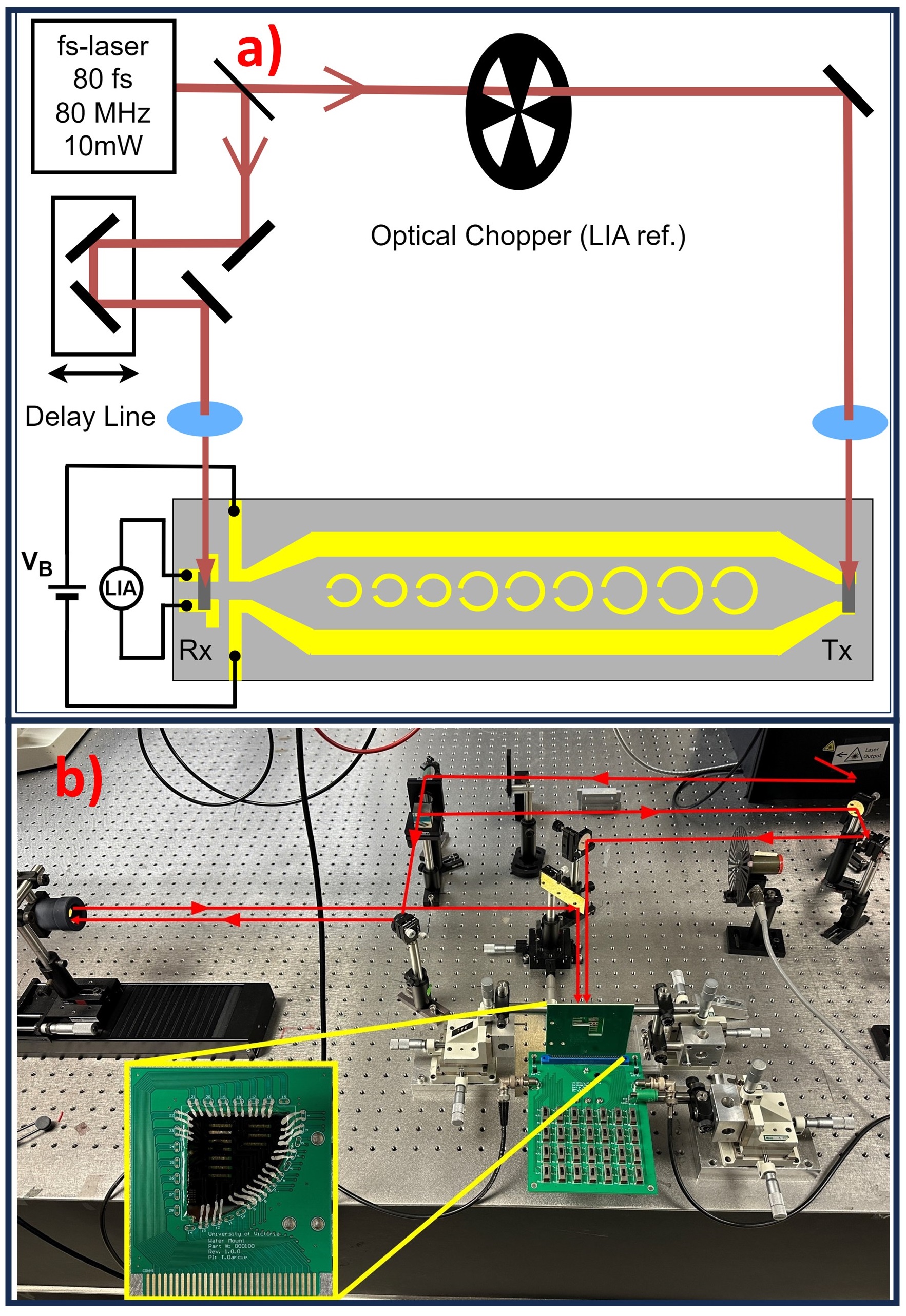}
\caption{Time-Domain Spectrometer. a) The transmitter path is optically chopped, and the receiver path passes through optical delay line. The THz-bandwidth pulse is generated at the DC-biased (VB) Tx LT-GaAs photoconducive switch, which then propagates to the Rx LT-GaAs photoconductive switch for detection by the lock-in amplifier (LIA) \cite{smith2021characterization}. b) The set-up used to do the experiment. The inset is a wafer on which Fig. 2 is fabricated. }
\label{fig:experiment_setup}
\end{figure}

\section{Experiment}

Figure \ref{fig:experiment_setup} illustrates the modified THz Time-Domain Spectrometer used to characterize the SRR BSF shown in Fig. \ref{fig:9srr_pic} at THz frequencies \cite{smith2019demonstration}. The procedure of the experiment is similar to \cite{smith2021characterization}. An 80 fs laser beam with 80 MHz frequency and 20 mW power divided into two paths, one is directed towards the transmitter and the other to the receiver. The transmitter beam path is optically chopped for the lock-in amplifier (LIA). After chopping, the optical beam is focused onto the DC-biased (VB = 25V) LT-GaAs photoconductive switch which generates a THz-bandwidth pulse propagating toward the receiver. To minimize radiation during excitation, the gap between the CPS conductors is initially made small. The CPS then tapers (from $S = 10~ \upmu$m and $W= 10~ \upmu$m) to a low-loss configuration ($S = 70~ \upmu$m and $W= 45~ \upmu$m). In the receiver path, the laser pulses propagate through a mechanical delay line prior to being focused onto receiver LT-GaAs photoconductive switch, which is connected to a lock-in amplifier (LIA). Translation of the delay line and subsequent signal processing reconstructs the received THz pulse.

\section{Results and Discussion}

\begin{figure}[h!]
(a)

\centering 
  \includegraphics[width=2.5in]{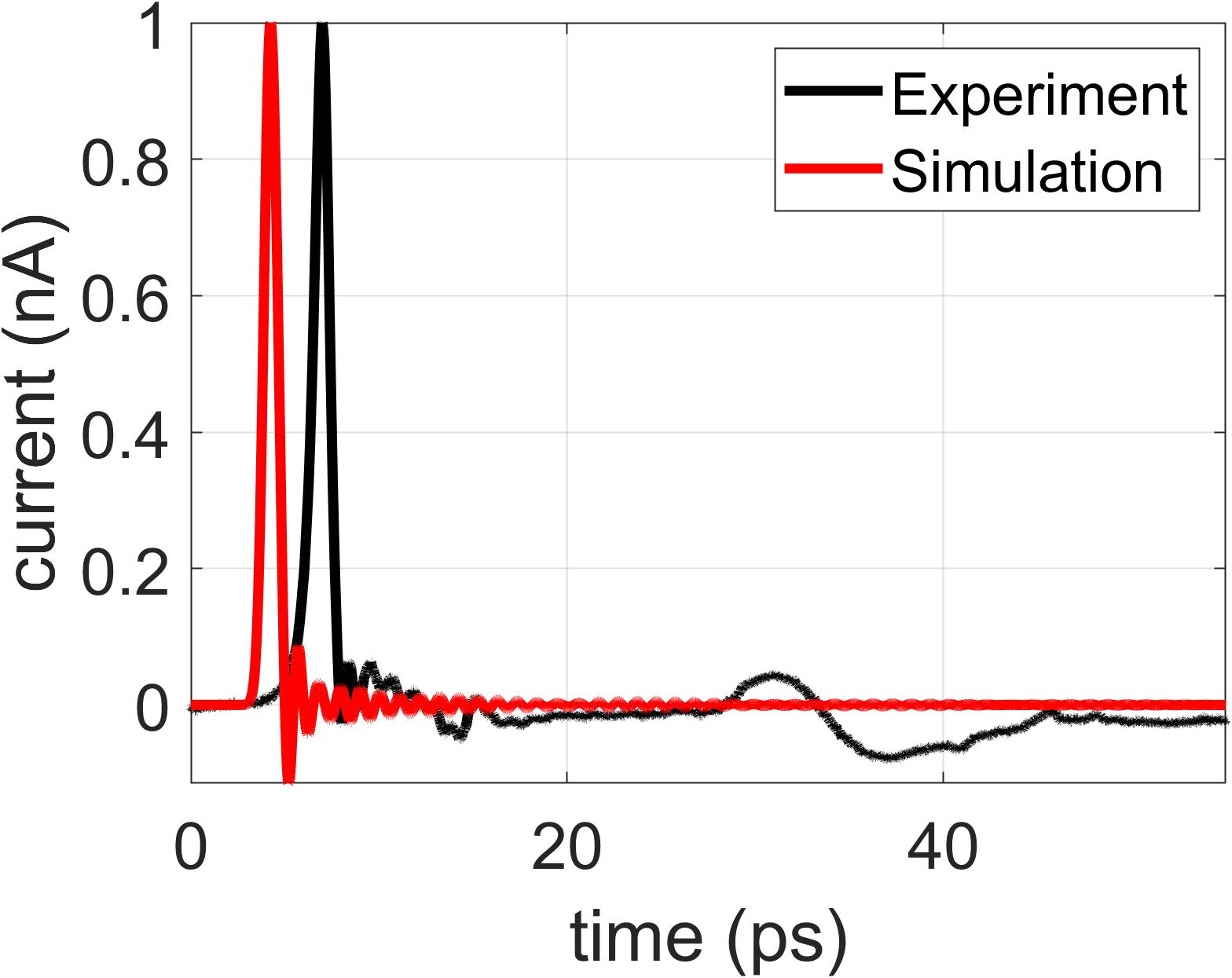}%
\begin{flushleft} 
(b)
\end{flushleft}
  \includegraphics[width=2.5in]{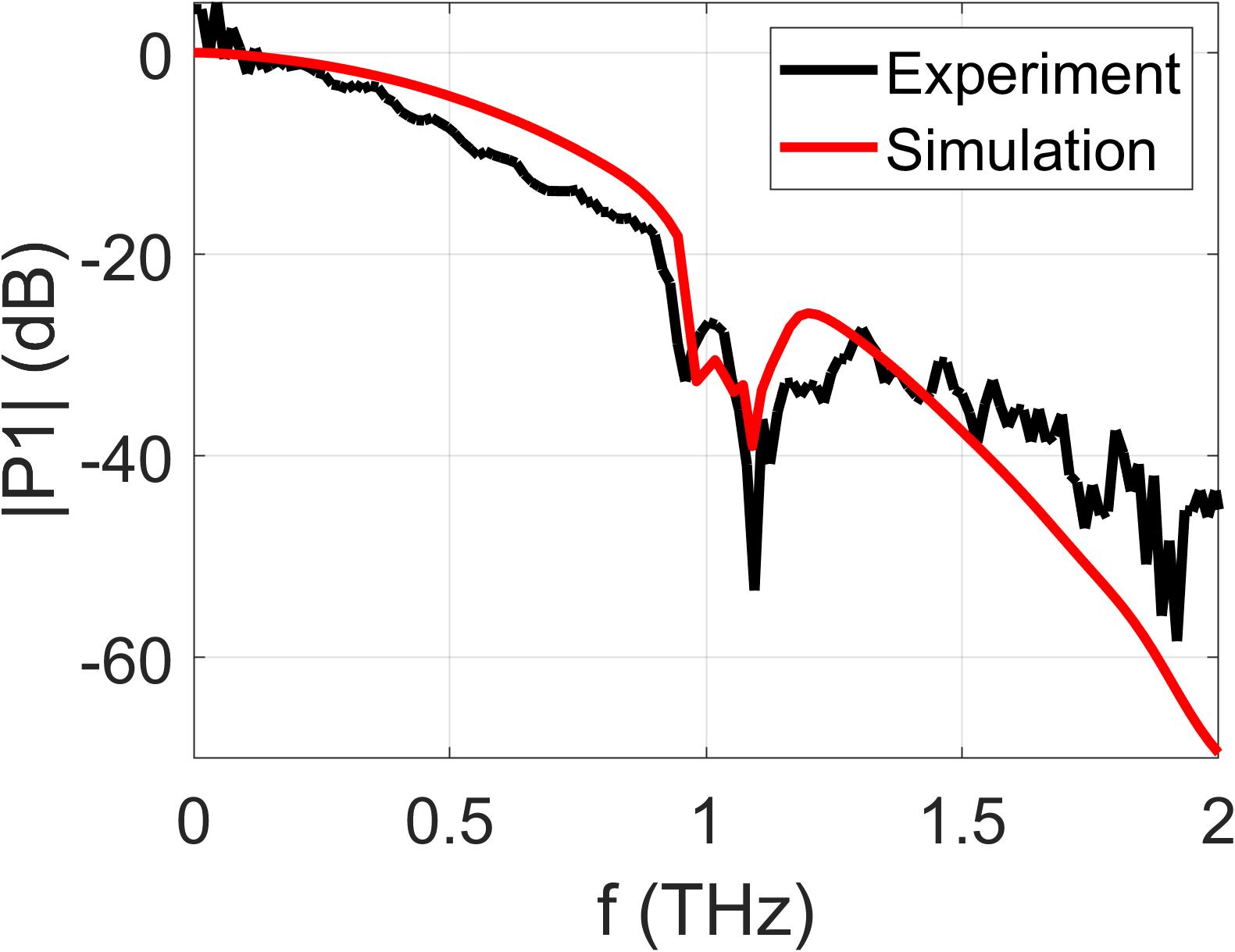}%

\caption{Experiment and simulation results. (a) time domain THz pulse at the receiver. (b) Fourier transform of the received THz pulse.}
\label{fig:experiment_results}
\end{figure}

The experimental result is plotted in Fig. \ref{fig:experiment_results} alongside a simulation trace. Figure \ref{fig:experiment_results}(a) plots the temporal response, and Fig. \ref{fig:experiment_results}(b) plots the spectral response which was obtained by applying the Discrete Fourier Transform to the temporal response. We find that the center of the stopband is located near 1 THz as designed, and the width of the stopband is about 0.36 THz which is larger than designed specification (via simulation). The deviation between experiment and simulation (bandwidth and rejection) arises from fabrication limitations and differences. The inner radii of each SRR group differs by 1 \textmu m. This is near the limit of our minimal photolithography feature size ($\approx$ 1 \textmu m). The impact of this limitation can be observed in the rounding of the SRR edges in Fig. \ref{fig:3srr_pic} which will result in unintended variations in SRR impedance. Regardless, in the experimental results, we do observe the desired bandstop characteristics thus demonstrating that THz stopband filters constructed of different radii SRRs can form a bandstop filter.

Future work will explore further optimization of the SRR configuration and SRR filter synthesis methodology.

\section{Conclusion}
This article reports fabrication, simulation, and experiment of a band stop filter with center frequency of 1.06 THz and a -3 dB bandwidth of 0.36 THz using SRRs of varying radii corresponding to different resonant frequencies. The presented experimental work serves as a proof-of-concept which demonstrates that SRRs are viable building blocks for subwavelength THz filters. 

\bibliography{bibliography.bib}
\bibliographystyle{IEEEtran}

\end{document}